\documentclass[%preprint,
amssymb,floatfix,floats,
aps,prl,preprintnumbers,noshowpacs,twocolumn,superscriptaddress]{revtex4-2}
\usepackage{booktabs}
\usepackage{color,graphicx}
\usepackage{adjustbox}
\usepackage[dvipsnames]{xcolor}
\usepackage{dcolumn}
\usepackage{upgreek}
\usepackage{bm}
\usepackage[hidelinks,colorlinks=true,allcolors=blue,citecolor=blue]{hyperref}
\usepackage{chngcntr}
\usepackage[thinspace,thinqspace]{SIunits}
\usepackage{amsfonts}
\usepackage[ulem=normalem]{changes}
\usepackage{orcidlink}
\usepackage{lineno}
\newcommand{\TN}{\ensuremath{T_\text{N}}}

\newcommand{\CCSF}{Cs$_2$Cu$_3$SnF$_{12}$}
\newcommand{\RCSF}{Rb$_2$Cu$_3$SnF$_{12}$}

\newcommand{\rs}[1]{\textcolor{black}{#1}}
\newcommand{\red}[1]{\rs{\sout{}}}

\begin{document} %\preprint{AIP/123-QED} 
\title{Discovery of BKT correlations in the quantum kagome compound \CCSF}

\author{M. S. Grbić\,\orcidlink{0000-0002-2542-2192}}
\email{corresponding author: mgrbic.phy@pmf.hr}
\affiliation{Department of Physics, Faculty of Science, University of Zagreb, Bijenička cesta 32, 10000 Zagreb, Croatia}

\author{I. Jakovac\,\orcidlink{0009-0005-0393-3601}}
\affiliation{Department of Physics, Faculty of Science, University of Zagreb, Bijenička cesta 32, 10000 Zagreb, Croatia}

\author{I. Kupčić\,\orcidlink{0009-0008-2102-0285}}
\affiliation{Department of Physics, Faculty of Science, University of Zagreb, Bijenička cesta 32, 10000 Zagreb, Croatia}

\author{H. Tanaka\,\orcidlink{0000-0001-8235-8783}}
\altaffiliation[Current address: ]{Center for Entrepreneurship Education, Institute of Science Tokyo, Midori-ku, Yokohama 226-8502, Japan}
\affiliation{Department of Physics, Institute of Science Tokyo, Meguro-ku, Tokyo 152-8551, Japan}

\author{M. Horvati{\'c}\,\orcidlink{0000-0001-7161-0488}}
\affiliation{Laboratoire National des Champs Magn\'{e}tiques Intenses, LNCMI-CNRS (UPR3228), EMFL, \\ Université Grenoble Alpes, Université Toulouse, INSA-T, 38042 Grenoble Cedex 9, France}

%\maketitle
\date{\today} 
\begin{abstract}
	We investigate the microscopic properties of the kagome compound \CCSF\ using $^{63,65}$Cu nuclear quadrupolar resonance (NQR). Analysis of the local hyperfine fields below the Néel temperature $\TN = 20$~K indicates a spin structure consistent with $P2_1/n$ symmetry of negative vector chirality. Measurements of the spin–lattice relaxation rate $T_1^{-1}$ reveal signatures of a gapless ground state and two-dimensional Berezinskii–Kosterlitz–Thouless (BKT)–type correlations above $\TN$, extending over a broad temperature range of approximately 130 K in zero magnetic field. Within the same temperature range, the observed increase in the NQR linewidth is consistent with short-range chiral order recently identified by neutron scattering. Our results establish \CCSF\	 as a unique quantum kagome system exhibiting BKT behavior.
\end{abstract}
\maketitle
%\\ \\
Frustrated quantum magnets, such as the spin $S$\,$=$\,$1/2$ triangular-lattice and the kagome-lattice Heisenberg antiferromagnets, have been extensively studied because a large degeneracy of their ground state and low-dimensional enhancement of quantum fluctuations allow exotic phenomena to emerge. However, these same features make it challenging to experimentally determine the exact nature of the ground state~\cite{Wen2019}, particularly if single crystals are unavailable. Despite these difficulties, significant progress has been achieved, and various quantum spin liquids, chiral orders and fractionalized excitations have been reported~\cite{Broholm2020}. Surprisingly, notwithstanding the large body of work on two-dimensional (2D) systems, only a single report to date has identified 2D XY correlations of Berezinskii-Kosterlitz-Thouless (BKT) type in a frustrated lattice \cite{Hu2020}.
One strategy of studying frustrated systems is to lift the ground-state degeneracy by controlled tuning - e.g. with chemical pressure~\cite{Hwang,Chatterjee,Hering2022} or external strain~\cite{Wang23} - thereby stabilizing a particular phase, that can be studied in detail.\\
\indent In the family of compounds A$_2$Cu$_3$SnF$_{12}$ (A = Rb, Cs) ~\cite{Ono2009,Katayama}, $S = 1/2$ is borne by the copper Cu$^{2+}$ ion enclosed in an F$_6$ octahedron, and the exchange interactions, mediated by the Cu--F--Cu bonds, form a kagome lattice. Cs and Rb lie outside the kagome plane and control the coupling between the layers. The two limiting cases, \RCSF\ (RCSF) and \CCSF\ (CCSF), show drastically different ground states. RCSF hosts a valence bond solid (VBS) singlet ground state with a particular ``pinwheel'' pattern~\cite{MatanNP,Grbic2013} formed by 4 different Heisenberg couplings --  $J_1 = 234$~K, $J_2 = 0.95 J_1$, $J_3 = 0.85J_1$, and $J_4 = 0.55J_1$. The large Dzyaloshinskii-Moriya (DM) interaction $d_i \approx 0.18J_i$ ($i=1$-$4$) on each bond defines the dispersion of the energy excitations and keeps the spin polarization within the kagome plane. CCSF is closer to the isotropic kagome~\cite{Ono2014} with $J_1 = 154$~K, $J_2 = J_1$, $J_3 = 0.84J_1$, $J_4 = 0.7J_1$, where frustration is higher. Although the interlayer coupling is weak ($J’/J_1 \le 10^{-3}$), a larger DM interaction of opposite sign, $d_i\approx -0.29J_i$, is believed to push the system~\cite{Ono2009,Cepas,Ferrari} to order below $\TN \approx 20$~K. In both systems, a perfect kagome lattice at room temperature on cooling goes through a structural transition that results with heterogeneous $J$ values. In CCSF, a neutron scattering (NS) study~\cite{Matan2019} has shown that the rhombohedral crystal structure (R\={3}m) reduces to the monoclinic one ($P2_1/n$)  below $T_t=186~$K. The same work found that magnetic structure corresponds to the $P2_1'/n'$ symmetry with positive chirality (i.e., all-in-all-out structure). Ground state excitations were shown to display a  spinon continuum~\cite{Saito2022,Matan2022} with negative renormalization~\cite{Ono2014,Kogure}, indicative of strong quantum effects driven by frustration and low $S$ value. 
Availability of large single crystals and unconventional physics makes CCSF a model for distorted-kagome system, for both experimental and theoretical studies. \\
\indent In this Letter we report on the discovery of 2D XY correlations of the Berezinskii-Kosterlitz-Thouless (BKT) type in the $S=1/2$ kagome compound CCSF, persisting in a broad temperature range. Its appearance is accompanied by 2D short-range order, connected to a chiral state recently observed by NS. We use $^{63,65}$Cu nuclear quadrupolar resonance (NQR) as a local experimental probe. In addition, by analyzing zero-field nuclear magnetic resonance (zf-NMR) spectra, we revise the exact spin structure of the long-range order. These results provide a very stringent framework for a theoretical description of the ground state realized in CCSF, and of 2D XY correlations in kagome compounds in general.\\
\begin{figure}
	\centering \vspace{0in}
	\includegraphics[width=\columnwidth]{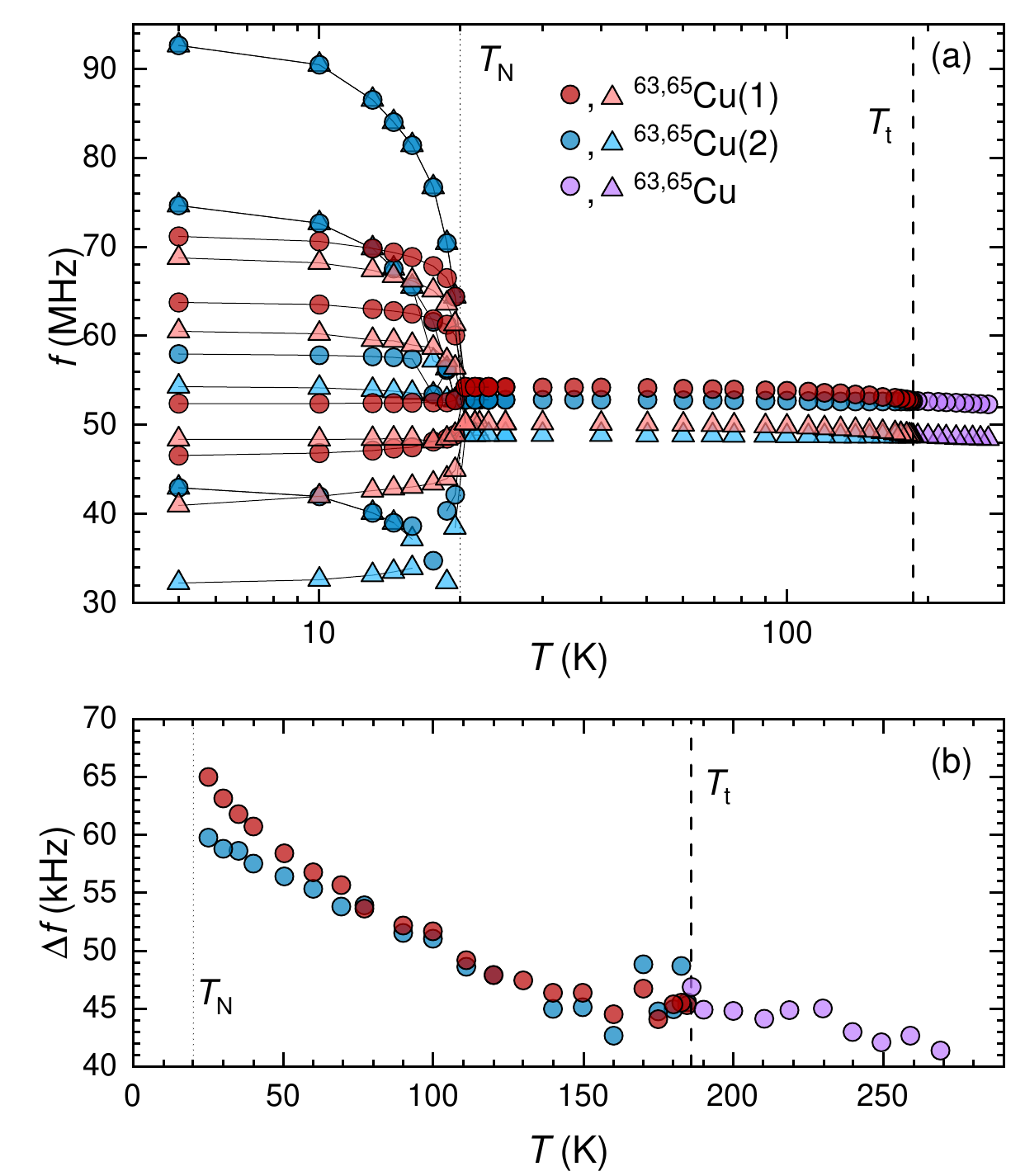}
	\caption{(a) NQR line is split below the structural transition temperature $T_t=186$~K. Below $\TN = 20$~K magnetic order onsets and the spectrum is further split. Circles and triangles mark the signals of $^{63,65}$Cu isotopes. Purple marks the signal of a single crystallographic site above $T_t$, while red and blue mark the signals of Cu(1) and Cu(2) sites, respectively. Lines are guides for the eyes. (b) $T$-dependence of NQR linewidths of Cu(1,2) sites show an increase below $\sim 150$~K. Note that temperature scales in (a) and (b) are different.}
	\label{fig1}
\end{figure}
\indent We first present the NQR spectral data. When a copper nucleus (spin $I=3/2$) is at a locally non-cubic site, it shows two NQR lines for each non-equivalent position at zero magnetic field. These correspond to the two isotopes: $^{63}$Cu ($^{65}$Cu) with gyromagnetic constants $^{63}\gamma=11.285$~MHz/T ($^{65}\gamma=12.089$~MHz/T), abundances of 69\% (31\%), respectively, and a ratio of quadrupolar moments $^{63}Q / ^{65}Q =1.0806$. The NQR frequency is defined by the nuclear quadrupolar Hamiltonian:
\begin{equation}
\mathcal{H}_Q = \frac{h\nu_Q}{6}\left[3I_z^2-I^2+\frac{\eta_Q(I_+^2 + I_-^2)}{2}\right],
\end{equation}
where $	\nu_Q = \frac{3eQ}{2I(2I-1)}V_{zz}$, $V_{zz}$ is the principal value of the local electric field gradient (EFG) tensor, $e$ is the elementary charge, $I_z$ and $I_+$($I_-$) are the standard $z$ projection and raising (lowering) operators of the nuclear spin, and $\eta_Q=(V_{yy}-V_{xx})/V_{zz}$ is the EFG asymmetry parameter. At 290~K there is one line per isotope (Figs.~\ref{fig1}(a) and S1) with a Lorentzian shape, with the frequency corresponding to $\nu_\text{NQR} = \nu_Q \sqrt{1+\eta^2/3}$. Typical FWHM linewidth of $\Delta f \approx43$~kHz shows the high quality of single crystals (Fig.~\ref{fig1}(b)). As the system is cooled across the structural transition at $T_t = 186$~K, NQR signal of a single Cu site splits into two: Cu(1) and Cu(2) (Figs.~\ref{fig1}(a) and S1). At low temperatures,  CuF$_6$  octahedra of Cu(1,2) sites are tilted by $\alpha_{1,2}=17.8^\circ (16^\circ)$, respectively. The angles $\alpha_{1,2}$ are defined by the average planes passing through the four nearest-neighbor F$^-$ ions that surround Cu, with respect to the kagome plane. These determine the orientation of the $d_{x^2-y^2}$ orbital and the orientation of the principal axes $\textbf{e}_\text{EFG}(1,2)$ corresponding to the principal value $V_{zz}$. NQR signal is assigned to a specific Cu site by comparing the signal intensities, which equal the structural occupancy of Cu sites in the unit cell $N[$Cu$(1)]$\,:\,$N[$Cu$(2)]$. The intensity ratio of $1$\,:\,$2$ (Fig. S1) is consistent with the expected one from the crystal structure of Ref.~[\onlinecite{Matan2019}]. As the system is cooled below $\approx 150$~K, the linewidth $\Delta f$ of both sites starts to increase (Fig.~\ref{fig1}(b)), reaching $60-65$~kHz just above \TN. This increase exactly coincides with the recently found onset of chiral state in CCSF~\cite{Matan2022}, indicating that the $\Delta f(T)$ depicts the formation of short-range magnetic order, expected to emerge~\cite{Corti,Sengupta2003} below $T\sim J$ (in our case $J_1$). A different chiral short-range 2D order has previously been observed in a classical kagome system~\cite{Schweika}. When CCSF is cooled below \TN, the N\' eel phase onsets, and a Zeeman term $H_M = -\gamma \hbar \textbf{B}_h \textbf{I}$ is then added to the nuclear Hamiltonian, where $\textbf{B}_h$ is the local hyperfine field. The shape of the spectrum, i.e. eigenvalues of the total Hamiltonian $H = H_M+H_Q$, is defined by the field size $B_\text{h}=|\textbf{B}_h|$, the angles $(\vartheta, \varphi)$ of $\textbf{B}_h$ with respect to $\textbf{e}_\text{EFG}$, and the values of $\nu_Q$ and $\eta_Q$. To determine $B_\text{h}$, each zf-NMR signal needs to be matched to a specific Cu site. Fortunately, the five mentioned parameters are reduced to two ($B_\text{h}$ and $\vartheta$) as NQR frequency $\nu_\text{NQR}$ is measured and $\eta_Q \approx 0$ is determined in a separate experiment (Fig.~S3). An additional point-charge calculation (see Supplemental Material) confirmed that $\eta_Q \approx 0$ and that the angle between $\textbf{e}_\text{EFG}$ and the kagome plane is within $2^\circ$ of $\alpha_{1,2}$. Since $\eta_Q  \approx 0$, the EFG tensor is axial and $\varphi$ becomes irrelevant for our calculation. The main difficulty in the spectral analysis then lies in the assignment of different lines in the NMR spectrum. The correct assignment was recognized by a successful fit of exactly diagonalized Hamiltonian $H$ to the complete spectra of both isotopes taken at different temperatures, leading to the fit parameters $B_h(i)$ and $\vartheta(i)$, where $i=1,2$ denotes the Cu sites. In Fig.~\ref{fig2}(b) (and Fig.~S4) we show the two sets of $B_h$ that account for the entire spectra. Surprisingly, the $B_h$ values are drastically different, with $B_h(1)\approx 0.5 B_h(2)$. Since the NS determined~\cite{Matan2019} that $|\textbf{m}(1)|\approx|\textbf{m}(2)|$, these $B_h$ values can only be explained by the magnetic moments with a distinct orientation with respect to the hyperfine coupling tensor $\textbf{A}$. Magnitude of $B_h$ is set by $\textbf{B}_h = \textbf{A}\cdot \textbf{m}$, where the typical values of the $\textbf{A}$ for the Cu nucleus in CuF$_6$~\cite{Itoh,Grbic2013} equal $A_{\parallel} \approx -18$~T/$\mu_B$ with a large anisotropy $A_{\parallel}/A_{\perp} \sim 10$. Here, ``$\parallel$'' denotes the principal axis of the $\textbf{A}$ tensor, typically parallel to $\textbf{e}_\text{EFG}$. Therefore, $|B_h(1)|\approx1$~T is realized if the moment $\textbf{m}(1)$ is oriented close to $A_\perp$. To generate $B_h(2) \approx 2B_h(1)$, $\textbf{m}(2)$ must form a much larger angle $\delta_\text{A}$ with respect to $A_\perp$, than $\textbf{m}(1)$.  The \textbf{all-in-all-out} spin structure that best fits the NS data is not consistent with these observations. However, it is important to note that the weak magnetic scattering intensity prevented NS to exclude any of the remaining three spin arrangements altogether. Therefore, the current  data provide the complementary information required to identify the correct structure. To this end, we calculate the angles $\delta_\text{A}(1,2)$ of the symmetry-allowed spin structures found by NS for Cu(1,2) sites and show them in Table~\ref{table1}. Only for the spin structure No.~4, having $P2_1/n$ symmetry and a negative vector chirality (-), we find that $\delta_\text{A}(2)> \delta_\text{A}(1)$ and present values that are consistent with the observed $B_h$ values. 
Interestingly, this structure is a predicted~\cite{Essafi} ground state for an isotropic kagome lattice with $d_z<0$ (as is the case for CCSF).
\begin{figure}
	\centering \vspace{0in}
	\includegraphics[width=\columnwidth]{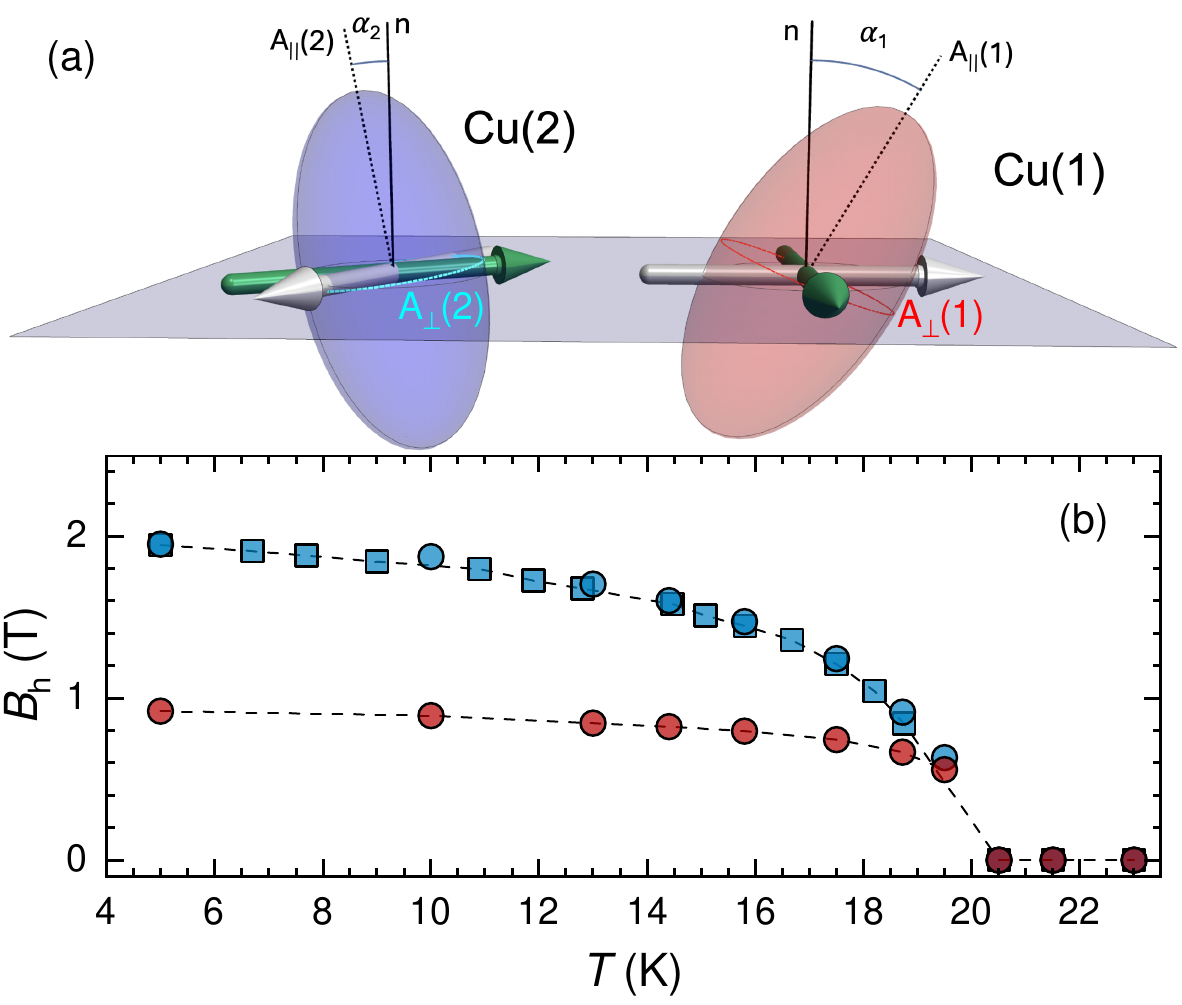}
	\caption{(a) Depiction of the hyperfine coupling (EFG) tensor tilting for angles $\alpha_{1,2}$ with respect to the kagome plane for Cu(1) and Cu(2) sites in red and blue, respectively. Tilting angles are emphasized for clarity.  Principal axes $A_\parallel$ ($\textbf{e}_\text{EFG}$) are marked with dashed black lines, kagome plane normal $\textbf{n}$ with full lines, $A_\perp$ component in cyan and red dashed circles. 3D arrows show the spin orientation on the Cu(1,2) sites determined by Ref.~\cite{Matan2019} (gray) and this work (green). See also Fig.~S5. (b) Fitting of spectra below \TN\ of Fig.~\ref{fig1}(a) results with local magnetic fields on Cu(1) (red circles) and Cu(2) (blue circles). Blue squares are additional points for Cu(2) obtained by measuring only the highest-frequency line and fixing the fitting parameters obtained from the spectra.}\label{fig2}
\end{figure}
\begin{table}[t]
	\caption{\label{table1} Angles $\delta_\text{A}(1,2)$ between the magnetic moment \textbf{m}(1,2) and $A_\perp$ component of the Cu(1,2) sites, for different spin structures proposed by neutron scattering~\cite{Matan2019}. Only structure No.~4 explains the zf-NMR data of this study.}
	\centering
	\begin{tabular}{c c c c }
		\hline \hline
		No.  & {Magnetic structure }
		&  \multicolumn{2}{c}{angle $\delta_\text{A}$ for site} \\
		
		&(vector chirality)&Cu(1) & Cu(2)\\
		\hline
		1 & $P2_1’/n’ (+)$ & $16^\circ$& $16^\circ$  \\
		2 & $P2_1’/n’ ~(-)$ & $16^\circ$ & $7.8^\circ$ \\
		3 & $P2_1/n ~(+)$ & $6.8^\circ$ & $0.1^\circ$\\
		4 & $\mathbf{P2_1/n ~(-)}$ & \textbf{6.8}$^\circ$& \textbf{13.9}$^\circ$\\
		\hline
		\hline
	\end{tabular}
\end{table}\\
\indent To study the spin dynamics, we have  measured the spin-spin ($T_2 ^{-1}$) and spin-lattice $(T_1 ^{-1})$ relaxation rates. $T_2 $ was determined by measuring the spin echo decay using a standard Hahn sequence ($\pi/2-\tau-\pi$) and fitting the data to $M(t)=M_0 e^{-t/T_2}$. $T_2 ^{-1}(T)$ dependence, shown in Fig.~\ref{fig3}(a), indicate the increase of local field fluctuations, from 270~K to low temperatures. As we approach \TN, below $\sim 25-30$~K, critical fluctuations of the 3D order start to build up and $T_2 ^{-1}$ of both sites diverge, reaching a maximum value at \TN. At lower temperatures the fluctuations are strongly reduced and $T_2$ becomes longer – saturating below 5~K to a value of 25~$\mu$s for Cu(2) site. One can see that below 120-150~K ($\approx J_1$) the two Cu sites present different $T_2 ^{-1}(T)$ behavior, indicating that some other fluctuation mechanism sets in. To further elucidate its origin, we focus on the $T_1$ data presented in Fig.~\ref{fig3}(b). $T_1$ time is a sensitive probe of intrinsic low-energy fluctuations of the system and is connected to the transverse spin correlation length $\xi$ as $1/T_1 \propto \xi^{z-\eta}$, where $z$ and $\eta$ are the dynamic and static critical exponents~\cite{Hohenberg}. By comparing $T_1$ data to $\xi$ of different systems we can identify the nature of underlying excitations. We have measured $T_1 ^{-1} (T)$ in detail on the Cu(2) site, and at several temperatures on Cu(1) site to verify it shows the same behavior. Relaxation curves were measured using a saturation-pulse sequence and fitted to the expression for $I=3/2$: $M(t) = M_0(1-e^{-(3t/T_1)^\beta})$ where $\beta$ is the stretch exponent. All the curves showed a single-component ($\beta \approx 1$) relaxation down to \TN.
\begin{figure}
	\centering \vspace{0in}
	\includegraphics[width=\columnwidth]{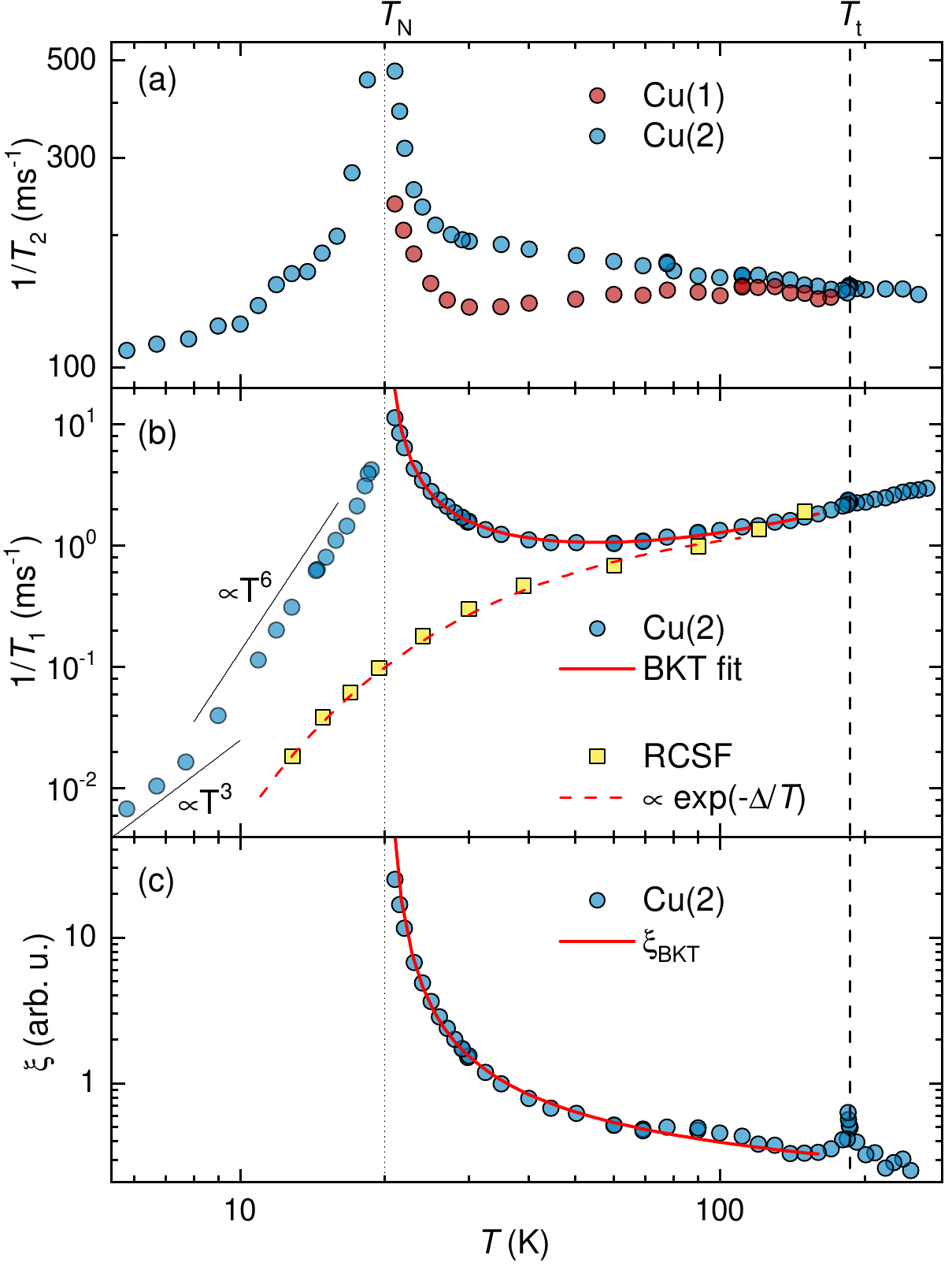}
	\caption{(a) $T_2 ^{-1}$ temperature dependence of both Cu sites. Below $\TN = 20$~K only the Cu(2) site was measured because of lower intensity of the Cu(1) site. Vertical dotted (dashed) line marks the magnetic (structural) transition temperature \TN\ ($T_t$). (b) $T_1 ^{-1}$ temperature dependence of Cu(2) site. Only a small increase is seen at $T_t$. Below $150$~K $T_1 ^{-1}$ starts to monotonously increase down to \TN. Full red line is a fit $T_1^{-1}  \propto\xi_\text{BKT}(T)^{z-\eta} + a_1 T+a_2 T^2$ (see main text). Below \TN, $T_1 ^{-1}$ drops first rapidly ($\propto T^6$) and below 9~K $\propto T^3$ indicating a gapless ground state. For comparison, we also show relaxation of Cu in RCSF which has a singlet ground state and shows no BKT divergence (yellow squares), but a gapped relaxation $T_1^{-1}  \propto \exp(-\Delta/T)$ (red dashed line). (c)~$\xi_\text{BKT}(T)$ calculated from (b). Full red line is the fit to $\xi_0\exp(0.5 \pi ⁄\sqrt{T⁄T_\text{BKT} -1})$.}\label{fig3}
\end{figure}
Generally, in antiferromagnetic insulators it is expected~\cite{Moriya} that $T_1 ^{-1}$ is temperature independent in the paramagnetic state. However, in CCSF,  $T_1 ^{-1}(T)$ dependence shows a clear quasi-linear background already below 300~K. To verify its origin, we have measured $T_1^{-1}$ up to 675~K  (Fig.~S6) and find that $T_1 ^{-1}$(T) develops also a $\propto T^2$ dependence above 400~K. From this we conclude that the background is due to a two-phonon absorption processes~\cite{Abragam} with a linear tail at low temperatures. We find the same background in the ``pinwheel'' VBS compound RCSF [(Fig.~\ref{fig3}(b)] which has akin crystal structure -- as expected for  the phononic origin. In CCSF, $T_1 ^{-1}$ rapidly decreases below \TN\ as the magnetic order develops; first with $T_1 ^{-1} \propto T^6$, and below 9~K as $\propto T^3$. The former behavior is quite common in the AF ordered states~\cite{Mayaffre2000} and corresponds to a higher-order relaxation process~\cite{Beeman}, while the later behavior probably indicates some saturation effect, as in Ref.~\cite{Jeong2017}. Both of these trends are power-laws, and show a vanishing energy scale (consistent with a small energy gap) at low temperatures. This is in-line with the NS observation of quantum spin liquid fluctuations in the ordered state~\cite{Saito2022}. Above \TN, we observe a large divergence, where $T_1 ^{-1}$ is increased by more than  an order of magnitude. In contrast, there is only a small increase from the structural transition at $T_t$. The enhanced $T_1 ^{-1}$, showing a build-up of spin fluctuations, can be traced quite high in temperature – all the way up to $150$~K ($\approx 7 \TN$). The magnitude and extent of this enhancement reveals a strong 2D character of the spin dynamics, as 3D critical excitations are typically confined within $\lesssim 10 \%$ above \TN. This argument is supported by the fact that in NQR, $T_2 ^{-1}$ ($T_1 ^{-1}$) probe the spin dynamics parallel (perpendicular) to $\textbf{e}_\text{EFG}$, respectively. 3D critical fluctuations should then emerge in $T_2 ^{-1}$ close to \TN, while  2D fluctuations (almost) perpendicular to $\textbf{e}_\text{EFG}$ are visible dominantly in $T_1 ^{-1}$. And indeed this is the case. We treat $T_1 ^{-1} (T)$ using several models (Fig.~S6): 3D and 2D Heisenberg antiferromagnet, and 2D BKT  correlation length of an easy-plane antiferromagnet $\xi_\text{BKT}(T) = \xi_0 \exp(0.5 \pi⁄\sqrt{T⁄T_\text{BKT} -1})$. The phononic background is accounted for using a second-order polynomial in $T$ without a constant term. We find that the fit using 2D BKT \textit{exactly} accounts for the observed increase, with $\xi_0 = 0.186(4)$ and $T_\text{BKT} = 19.31(5)$~K. The $T_\text{BKT}$ temperature that is very close to \TN\ shows that the system is strongly 2D, with the interlayer coupling~\cite{Laflorencie2012,Yasuda2005} of $J’/J_\text{avg.} \approx10^{-4}$. $\xi_\text{BKT}(T)$ has a characteristic temperature dependence, which is why fitting the data using other models fails to reproduce the observed behavior. By isolating the $\xi_\text{BKT}(T)$ dependence (Fig.~\ref{fig3}(c)) from the $T_1 ^{-1}$ data we show that the enhancement in $T_1 ^{-1} (T)$, visible below $150$~K, indeed originates from 2D fluctuations, coinciding with the  the onset of 2D short-range chiral state~\cite{Matan2022} detected by linewidth $\Delta f(T)$.\\
\indent BKT correlations have so far been reported in a handfull of cases: for $S=1/2$ systems, in square~\cite{Suh,Opherden,Ranjith2026} and triangular~\cite{Hu2020} lattices, and for $S=1$ in a honeycomb~\cite{Heinrich} lattice. In each case, correlations were confined to a narrow temperature range, in contrast to CCSF. Surprisingly, despite numerous studies of kagome systems, BKT correlations were not reported to date. To the best of our knowledge, similar behavior was observed only in one other system - the Fe jarosite~\cite{Matan2006,Nishiyama}, where the spin of $S=5/2$ is rather in the classical limit. Interestingly, in that system as well, NS found unconventional XY chiral order~\cite{Grohol2005} above \TN, just like in CCSF. The lack of reports of BKT physics in kagome systems naturally imposes a question on the possible cause. Anisotropy which forces the spins into the kagome plane clearly plays an important role: in Fe-jarosite, DM interaction and/or single-ion anisotropy~\cite{Matan2006}, and in CCSF, a strong DM interaction $ d_z = -0.29 J$. This is also the reason why in CCSF BKT correlations are visible in zero magnetic field, and in such a wide temperature range~\cite{Herak2014,Cvitanic2018}. Large DM interaction is also present ($0.18J$) in the ``pinwheel'' compound RCSF, with a different sign, which has been shown to play an important role~\cite{Grbic2013}: the sign of the DM interaction is connected to the inherited chirality of the triangular lattice and  $d_z >0$ opens a gap that prevents the onset of order. This is why BKT correlations are not observed in the (gapped) ``pinwheel'' compound RCSF. However, this also indicates that there is a BKT-onset point between RCSF and CCSF that can be reached by doping. A magnetic quantum critical point~\cite{Katayama} has been found at $x=0.53$ in $($Rb$_{1-x}$Cs$_x )_2$Cu$_3$SnF$_{12}$, but there are no reports on underlying excitations of the system. The influence of anisotropy on BKT physics has been theoretically studied for several other cases~\cite{Kim,Li2020,Caci}, but these are not applicable to the present case.\\
\indent In summary, we found BKT correlations in the $S=1/2$ kagome compound \CCSF\ that persist in a broad temperature range. It onsets together with the 2D XY short range chiral order below $T\sim J_1\approx 150~$K. By analyzing the zero-field nuclear magnetic resonance spectra, we revise the exact spin structure of the long-range order in this compound. Lack of 2D XY correlations in the sister-compound \RCSF\ with a VBS ground state, and the availability of $($Rb$_{1-x}$Cs$_x )_2$Cu$_3$SnF$_{12}$ compound that continuously tunes between the two behaviors, shows a direct path for future theoretical and experimental studies on the onset of 2D XY correlations in kagome compounds in general.

\begin{acknowledgments}
	We acknowledge fruitful discussions with K.~Matan, T.~Ono and D.~Pajić. We acknowledge the help of M.~Novak in developing the high-temperature setup. M.S.G. and I.J. acknowledge the support of Unity Through Knowledge Fund (UKF Grant No. 20/15), Croatian Science Foundation (HRZZ) under the Project No. IP-2024-05-4586 and the support of project CeNIKS co-financed by the Croatian Government and the European Union through the European Regional Development Fund - Competitiveness and Cohesion Operational Programme (Grant No. KK.01.1.1.02.0013).
\end{acknowledgments}

%\bibliography{BibList}{} 
%\bibliographystyle{unsrtdin} % \vspace{0.5cm}
%\renewcommand{\bibsection}{\section*{References}}
%\bibliography{references}
%\bibliographystyle{naturemag}
%
%%%%%%%%%%%%%%%%%%%%%%%%%%%%%%%%%%%%%%%%%%%%%%%%%%%5

\end{document}